# Magnetic-field dependent vortex dynamics and critical currents in superconducting microwires with regular large-area perforation by pinholes

Dong Zhu[1], Ilya Charaev[1], Konstantin Ilin[2], Andreas Schilling[1*]

[1]Department of Physics, University of Zurich, Winterthurerstrasse 190, CH-8057 Zurich, Switzerland

[2]Institute of Micro- and Nanoelectronic Systems, Karlsruhe Institute of Technology (KIT), 76187, Karlsruhe, Germany

*We report on results of simulations and experiments of vortex states in superconducting micro-wires with periodic rectangular pinhole structures. The simulations have been performed by means of numerically solving the time-dependent Ginzburg-Landau (TDGL) equations. With increasing bias current and for different values of the external magnetic field applied normal to the structure plane, we observe at first a vortex free Meissner state, followed by a resistive vortex-flow mixed state and a state with a more complex vortex pattern. The resulting dependence of the critical current $I_c$ on magnetic field exhibits two plateaus with distinctly different vortex dynamics. Corresponding experimentally measured magnetic-field dependences of $I_c$ of WSi microwires with periodic pinhole structures and varying hole spacing confirmed the predictions of these simulations, showing two ranges of magnetic field with almost field-independent critical currents. The experimentally determined critical currents are larger for a smaller pinhole spacing, in agreement with the results of the TDGL simulations. The good agreement of the simulations with the experimental results provides a convenient strategy for the optimization of single-photon detectors with or without artificial and natural defects.*

*Corresponding author. *Email address:* schilling@physik.uzh.ch (A. Schilling)

# Introduction

Superconducting micrometer-wide single-photon detectors (SMSPD), which can be fabricated of a variety of materials such as NbN, MoSi, WSi and NbRe, have recently attracted significant attention [1-7]. The extension from nanowires to micrometer-wide wires and the corresponding expected alteration of the light detection mechanism at work offers many avenues for further exploration. For SMSPDs, the mechanism of light-quanta detection is likely based on vortex generation and motion. The energy released by normal-state electrons accompanying the vortex motion generates a voltage pulse that can be detected by an external circuit. Therefore, the vortex dynamics in microwires is a particularly important and valuable subject to study [1, 8, 9]. Recently, it has not only been experimentally shown [5] that dark counts of such detectors in a magnetic field can be suppressed effectively by perforating the microwires with a grid of pinholes, but also that the magnetic-field dependence of the critical current shows a peculiar plateau in low magnetic fields around zero. Therefore, artificially perforated pinhole structures or random defects created by energetic ion bombardment can drastically affect the photon-detection performance in superconducting micro- and nanowires [4, 10-12].

For superconducting thin films or micrometer and sub-micrometer wide strips without pinholes, single or multiple vortex motion has been investigated theoretically and computationally [9, 13, 14]. Vortex motion is directly causally related to the magnitude of the critical current as defined by the current-induced transition from the superconducting to a resistive state. In case of a single defect (one hole or an area of suppressed superconductivity), a qualitative analysis already allows for a good description and prediction of the behavior of the system. As reported by Vodolazov *et al.* [15], two mechanisms, i.e., the vortex entry via the edge of the microstrip, and the creation of vortex-anti-vortex pairs nucleating in the vicinity of the hole, determine the critical current in such structures in the presence of an applied current and in an external magnetic field. In case of many spatially distributed defects, a simple qualitative analysis of such complex systems is no longer possible, however, and, therefore, simulations on the basis of time dependent Ginzburg-Landau theory (TDGL) are required [16].

In this work, we computationally studied the vortex dynamics in superconducting microwires with three columns of pinhole chains with varying pinhole distance in a rectangular periodic arrangement. The TDGL simulations directly show the dynamics of the vortex state of the microwires in the presence of a bias current and an external magnetic field, with vortex-free Meissner states that are well distinguishable from resistive vortex-flow states appearing with increasing current and show a remarkable complexity of the vortex arrangement at large bias currents. Using a suitable realistic voltage threshold, our simulations allow us also to determine the critical current as a function of the magnetic field.

We find that the critical current in varying magnetic fields shows two distinct plateaus, regardless of the size of the pinhole spacing. The time-dependent vortex evolution near these two critical-current plateaus suggests that the presence of the pinholes enhances the vortex-free Meissner state in the first plateau and stabilizes the vortex-flow mixed state in the second plateau. To verify the results of these simulations, we have experimentally examined WSi microwires with corresponding pinhole structures for three different longitudinal pinhole spacings and have actually observed the two predicted critical-current plateaus in an external magnetic field. By comparing the results for the different pinhole spacings, we also found that smaller hole spacings correspond to higher critical currents, suggesting an influence on the vortex dynamics by the geometrical configuration of the pinholes and improving the current-carrying capability of the microwires.

## Simulation details

The TDGL simulation in this work was mainly performed using the python computational package py-TDGL presented in [17], which can solve vortex and phase dynamics in arbitrarily shaped 2D superconducting thin films, with an applied external magnetic field or bias current, or both. Py-TDGL is based on the generalized time-dependent Ginzburg-Landau equation, which has the dimensionless form,

$$\frac{u}{\sqrt{1+\gamma^2|\psi|^2}}\left(\frac{\partial}{\partial t} + i\,\varphi(\boldsymbol{r},t) + \frac{\gamma^2}{2}\frac{\partial|\psi|^2}{\partial t}\right)\psi(\boldsymbol{r},t) = (\epsilon - |\psi|^2)\psi(\boldsymbol{r},t) + \left(\nabla - i\,\boldsymbol{A}(\boldsymbol{r},t)\right)^2\psi(\boldsymbol{r},t), \qquad (1)$$

where $\psi(\boldsymbol{r},t) = |\psi|e^{i\theta}$ is the superconducting order parameter describing the superconducting condensate at position $\boldsymbol{r}$ and time $t$, and $\varphi(\boldsymbol{r},t)$ and $\boldsymbol{A}(\boldsymbol{r},t)$ are the Coulomb and the magnetic vector potentials in the superconducting film, respectively. The parameter $\epsilon$ is a real-valued number adjusting the local critical temperature of the film and is set here to 1 by default. The constant $u = \pi^4/14\zeta(3) \approx 5.79$ is the ratio of relaxation times for the amplitude and the phase of the order parameter in dirty superconductors, and $\gamma$ is a measure for the inelastic electron-phonon scattering strength which is set to 1 in our simulations [8, 9, 17].

The potential fields associated with the normal-current density satisfy the equation

$$\nabla^2 \varphi(\mathbf{r},t) = \nabla \cdot \mathbf{J}_s = \nabla \cdot \operatorname{Im} \psi^*(\nabla - i\mathbf{A}(\mathbf{r},t))\psi(\mathbf{r},t), \tag{2}$$

(where $\mathbf{J}_s$ is the supercurrent density), accounting for the fact that the total current in the film is divergence free. Together with appropriate boundary conditions, py-TDGL employs a finite element analysis to solve the time dependence of the order parameter, the supercurrent density, the normal-current density, the Coulomb potential and the magnetic-field distribution in superconducting thin films.

In these simulations, we can not only probe the time evolution of the order parameter $\psi(\mathbf{r},t)$ and directly observe the creation of magnetic vortices. By setting voltage probing points as shown in Fig. 1(a), we can also obtain the evolution of the voltage over time due to vortex flow and define a critical current using a suitably chosen voltage threshold. According to the py-TDGL simulation package where the G-L equations are dimensionless [17], we can convert the corresponding dimensionless times and voltages to physical units by multiplying them with the characteristic time $\tau_0 = \mu_0 \sigma \lambda^2$ and voltage $V_0 = \frac{4\xi^2 B_{c2}}{\mu_0 \sigma \lambda^2} = \frac{2\Phi_0}{\pi \mu_0 \sigma \lambda^2}$, respectively, where $\lambda$ is the magnetic penetration depth, $\xi$ the coherence length, $\mu_0$ the permeability of the vacuum, $\Phi_0$ the magnetic flux quantum, and $B_{c2}$ the upper-critical field. For a typical WSi microwire, the conductivity $\sigma = \frac{1}{R_s d} = 5 \times 10^5$ S/m, while the normal sheet resistance $R_s$ and the thickness $d$ of the strip are set to $R_s = 1000\ \Omega$ [5] and $d = 2$ nm, respectively, to account for the actual values in the subsequent experiments. The resulting characteristic time and voltage units are, with chosen $\lambda = 960$ nm (see Supplementary material), $\tau_0 = 0.58$ ps and $V_0 = 2.3$ mV, respectively.

# Results

For our numerical calculations we have initially modelled micro-line 1 (in the following abbreviated as "line-1"), with length $L = 3$ μm and width $w = 1$ μm, see Fig. 1(a). It contains pinholes with hole diameter $D = 100$ nm, arranged in a rectangular configuration with equal hole spacings $\Delta x = \Delta y = 200$ nm in the transverse and in the longitudinal directions, respectively. In our calculations, we used parameters of a typical WSi microwire with a thickness of 2 nm, with a London penetration depth $\lambda$ = 960 nm as obtained from the measured kinetic inductance in pulse-shape experiments on the devices as described in the Supplementary material, and varying coherence lengths of the order of $\xi \approx 13$ nm [5]. In Fig. 1(b), we show the resulting mixed state of this line in an external magnetic field of 80 mT and without bias current as an example. As expected

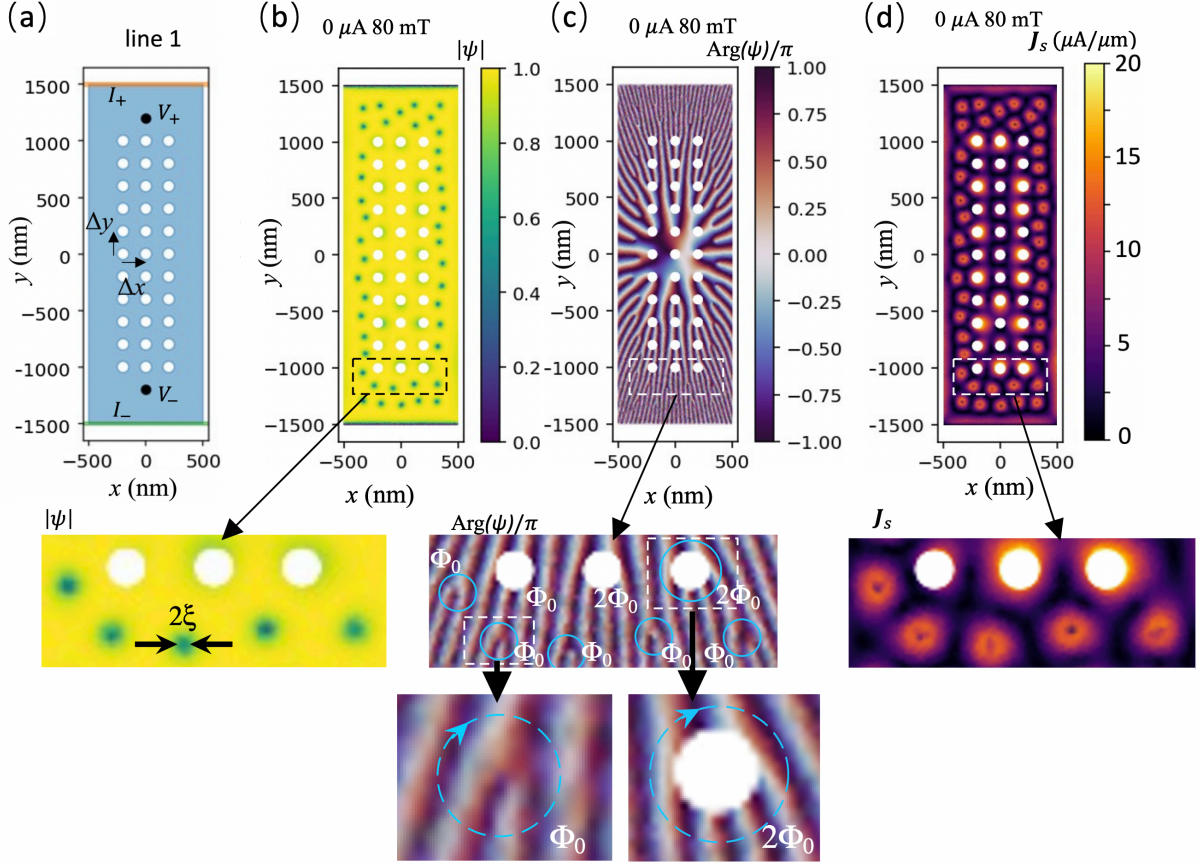

*Figure 1. (a) Design of the superconducting micro-line 1 perforated with pinholes, with a transverse width w = 1 µm and length L = 3 µm. The pinholes with diameter D = 100 nm are arranged in a rectangular arrangement, for line 1 at equal distances $\Delta x$ = 200 nm in the transverse and $\Delta y$ = 200 nm in the longitudinal direction, respectively; (b, c, d) the mixed state of micro-line 1 as an example, in an applied magnetic field of 80 mT and without bias current. The figures, from left to right, show (b) the amplitude $|\psi|$ of order parameter, (c) its phase, and (d) the corresponding supercurrent density $J_s$ per unit width, with circular screening currents forming the vortices. A magnification of the dashed rectangular areas is shown below these figures, indicating the size $\approx 2\xi$ of a vortex core in the left panel. A further magnification of the phase map (bottom panels) illustrates the bifurcation phenomenon for a single vortex carrying one flux quantum $\Phi_0$ (left bottom panel) and for a vortex with two flux quanta, trapped by a pinhole (right bottom panel), marked with blue dashed circles. The London magnetic penetration depth $\lambda$ and the coherent length $\xi$ for this simulation were chosen as 960 nm, and 13.6 nm, respectively.*

[18], quantized vortices appear in the sample, with the order parameter amplitude $|\psi|$ gradually decreasing at the centers (green circles with approximate diameter $2\xi$, see magnification in the lower left panel of Fig. 1). The phase of the order parameter $\text{Arg}(\psi)$ is shown in Fig. 1(c). The unpinned vortices are mainly distributed at the edge of the sample, and the resulting superconducting screening currents (red rings) can be clearly seen in Fig. 1(d) [19]. The red color along the edges of the line represents the Meissner shielding current. The overall phase-pattern shows a divergent distribution from the center outward, with a slight asymmetry due to non-uniform magnetization as a consequence of the non-uniform vortex distribution. Unlike for the freely moving vortices, the phase changes around certain pinholes are $4\pi$ (corresponding to two magnetic-flux quanta, $2\Phi_0$), such as for the top left and the bottom right two pinholes, among others, while the majority show a phase change of $2\pi$ as expected, carrying one magnetic-flux quantum $\Phi_0$. These phase changes manifest themselves in corresponding bifurcations in the phase map as displayed in the corresponding magnifications of Fig. 1(c) (lower panels).

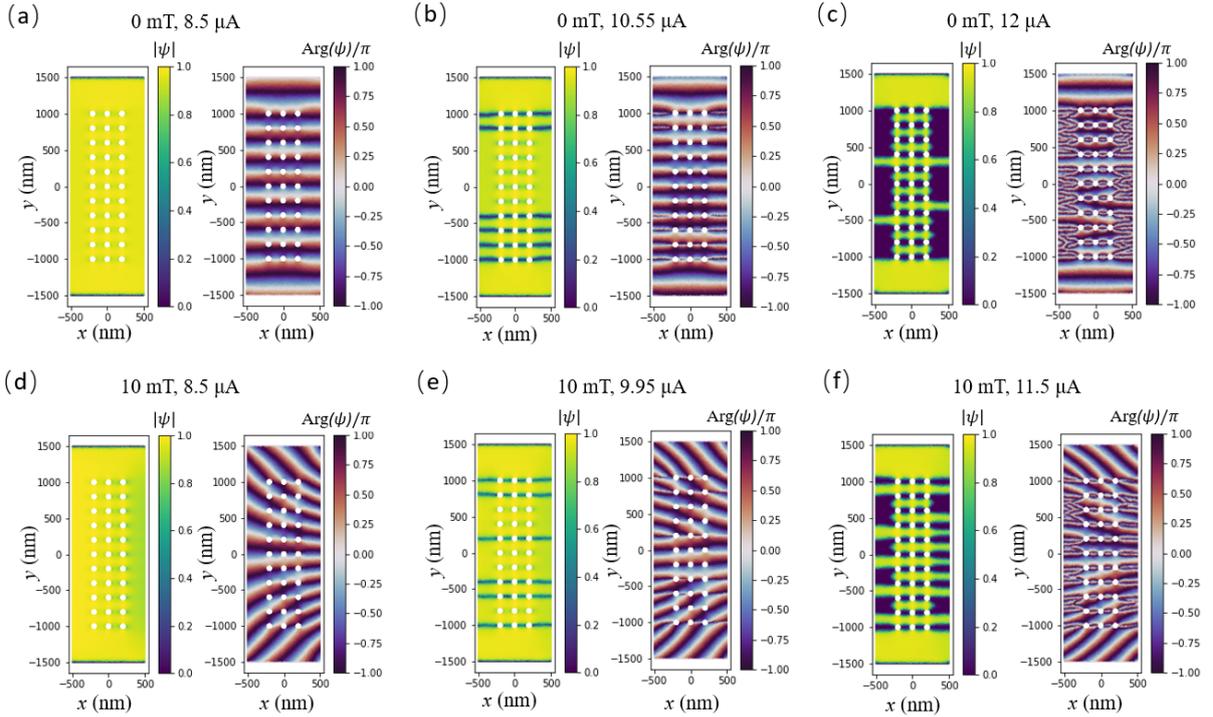

*Figure 2. Amplitude $|\psi|$ and phase $\text{Arg}(\psi)/\pi$ of line 1 at different bias currents as indicated on top of each pair of figures for two selected magnetic fields 0 mT (a-c) and 10 mT (d-f), with the bias current in the y-direction flowing from top to bottom, and the magnetic field directing out of the sample plane; (a, d) vortex-free Meissner states; (b, e) vortex-flow mixed states around the respective critical currents $I_c$, and (c, f) above $I_c$ in the resistive state, showing patchy normal-state regions in near the pinholes (see also the movies provided in the Supplementary material for details and the possibility to zoom into the graphs).*

Figure 2 shows the amplitudes and phases of the superconducting order parameter $\psi$ in the superconducting state of the line 1 at different bias currents and for two values of the external magnetic field of 0 and 10 mT. The same material parameters as in Fig. 1 ($\lambda$ = 960 nm and $\xi$ = 13.6 nm) were used for this simulation. With increasing current, the superconducting state appears to transform from a pure Meissner state to a state with a regular vortex-flow distribution slightly below the respective critical currents $I_c$, and eventually to a resistive state at even larger currents. To define these critical-current values, we have chosen a voltage-threshold criterion to be discussed in detail further below.

In zero magnetic field and below the critical current as shown in Fig. 2(a), the amplitude of the order parameter remains almost constant at 1, and the phase exhibits a plane-wave periodic variation reflecting the flow of the supercurrents. The absence of significant variations of the order parameter near the pinholes indicates that there is no vortex or vortex-pair generation at these sites. A slight suppression of the amplitude of order parameter between the pinholes reflects weak current crowding. In $B$ = 10 mT and below the respective critical current, Fig. 2(d), the amplitude of the order parameter is partly suppressed at the right half of the strip. The spatial variation of the phase along the $y$-axis shows a distinct gradient, which reflects the uneven distribution of the supercurrents, but the system still remains in pure Meissner state as no vortices appear [20]. This situation can be described qualitatively in following way. The suppression of the order parameter at the right edge of the strips is a consequence of the fact that the bias and Meissner currents are flowing in the same direction, thereby reducing the order parameter more than at the left edge where these current directions are opposite. A certain suppression of the order parameter is also seen between the middle and the right column of the pinholes and at the right side of the latter pinholes. Here, the local current density is increased due to current crowding.

In the state of vortex flow at currents near the respective critical values as shown in Fig. 2 (b, e), several continuous horizonal lines with strongly suppressed order parameter appear, connecting the pinholes. For example, in zero magnetic field, Fig. 2(b), the amplitude of the order parameter is suppressed across the full width of the strip in six rows of pinholes. In the other regions, the spatial distribution of the order parameter remains virtually unchanged as compared to Fig. 2(a) [21, 22]. In a magnetic field of 10 mT, Fig. 2(e), the phase of the order parameter shows a clear bifurcation phenomenon at certain pinhole sites, which is a typical indication that vortex lines are trapped at these positions.

At even larger bias currents beyond the critical current, a vortex-flow state appears that shows peculiar patterns as shown in Fig. 2(c, f). The order parameter is suppressed in such a manner that patchy normal-state regions form in proximity to the pinholes, resulting in a phase landscape that becomes exceedingly delicate and appears almost chaotic on both sides of the pinhole rows [9, 13].

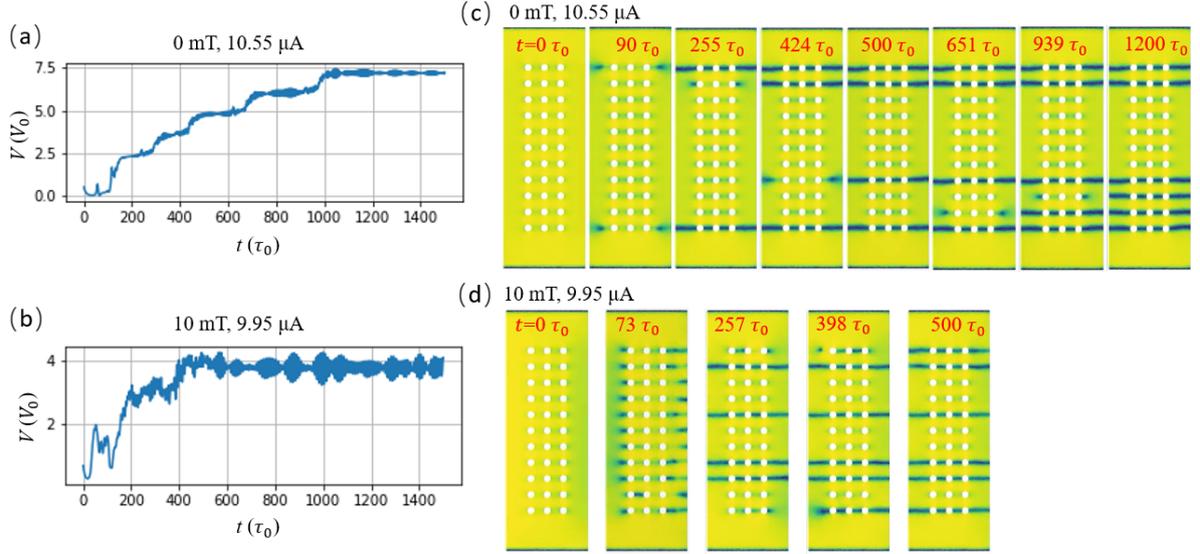

*Figure 3. (a, b) Evolution of the voltage across the line 1 over a 1500 $\tau_0 \approx 0.9$ ns time period for two sets of magnetic field and bias current around the respective critical current, assuming a coherence length $\xi = 13.6$ nm; (c) Time slices of the order-parameter amplitude at 0 mT and 10.55 µA; (d) Corresponding data for 10 mT and 9.95 µA.*

We now turn to the details of the time evolution of the voltage across the bridge according to our simulations of line 1. After applying a bias current of 10.55 µA at $t = 0$ in zero magnetic field, vortex-antivortex pairs are formed and move horizontally outwards from the pinholes to the edges within certain pinhole rows, accompanied by a voltage increase from zero due to the dissipative vortex motion. The generated voltage turns out to be proportional to the number of such rows showing vortex flow, and it therefore exhibits a stepwise increase over time as depicted in Fig. 3(a). This reflects the fact that the generation of vortices is discontinuous, and a transient saturation of the number of flowing vortices occurs at each voltage step. At the first step, the voltage rises by 2.5 voltage units, while the subsequent four steps correspond to 1.25 units, summing up to 7.5 units and 6 affected pinhole rows in the final state, as also shown in Fig. 2(b). The 1:1 correspondence of voltage and the number of pinhole rows exhibiting regular vortex flow can clearly been seen by inspecting the time sequence of the order parameter of line 1 in Fig. 3 (c), and in the detailed movies provided in the Supplementary material. The fact that vortices first appear at the top and the bottom of the microstrip is due to the current crowding at both ends where the uniform current meets the field of pinholes, and the region of a vortex-free Meissner state is kept as large as possible throughout the whole process.

The corresponding evolution of the voltage over time in a magnetic field of 10 mT with a bias current of 9.55 µA is shown in Fig. 3(b,d), along with the accompanying variations of the order parameter and the creation of line structures connecting pinholes within individual rows. The system starts already with a weak current-crowding effect due to the presence of the Meissner current. In contrast to the zero-field case, vortex flow is initially triggered in all pinhole rows immediately after turning on the bias current due to this current crowding, as it is shown for $t = 73\ \tau_0$ in Fig. 3(d), and the voltage rapidly increases to 2 voltage units before partially decaying to a lower value. Afterwards, as in the zero-field case, more and more pinhole-rows exhibit continuous vortex flow with increasing duration of the process, until a final state is reached with a stable voltage. The fact that the voltage (and therefore resistance) steps in zero magnetic field and in 10 mT are not equal stems for the fact that the order-parameter suppression within a pinhole row for the chosen different bias currents turns out to be different in 10 mT than in zero field. We are presenting a complete set of simulated $I$-$V$ data in the Supplementary material.

Since the dissipated power caused by vortex motion can drive a microwire from the superconducting to the normal state, a threshold voltage should be chosen to define the value of the critical current $I_c$ and its variation in an external magnetic field that can eventually be compared with experimental results. We note, however, that the evolution of the voltage as shown in Figs. 3 and the $I$-$V$ curves presented in the Supplementary material correspond to a case where Joule heat has no direct effect on the behavior of the electronic system, i.e, a situation with perfect thermal anchoring as it could be realized, for instance, by immersing the thin film in superfluid helium. In reality, the transition to the normal state can result in a local temperature increase within the film material that can ultimately lead to thermal latching (see Supplementary material), so that measured $V(t)$ and $I$-$V$ curves would deviate from those obtained by the present simulations. Nevertheless, the bias current leading to a suitably chosen low threshold voltage should still reflect the technical critical current $I_c$ that is measured in an actual experiment, above which possible thermal latching would occur.

We have chosen $4\ V_0$ ($\approx 9$ mV) as a threshold voltage, which corresponds to stable vortex flow across about half of the pinhole rows for line 1 in $B = 10$ mT. Figure 4 (a) shows the dependence of the critical current $I_c(B)$ on the magnetic field as calculated for four different coherence lengths, $\xi = 13$ nm, 13.4 nm, 13.6 nm, and 13.8 nm, respectively. In zero magnetic field, the simulated $I_c(0)$ decrease with increasing coherence length, and correspond to 11.0 µA, 10.68 µA, 10.53 µA and 10.4 µA, respectively. In a weak magnetic field up to 1-2 mT, the critical currents remain almost constant, i.e. independent of the applied magnetic field. A further increase of $B$ beyond 2 mT up to ≈ 9 mT results at first in a decrease of $I_c$, but it then becomes again comparably weakly dependent on magnetic field up to $B ≈$ 13 mT, thereby forming a second plateau-like feature in $I_c(B)$. The $I_c$ values at the second plateau also decrease with the increasing $ξ$, to 10.34 µA, 10.08 µA, 9.95 µA, and 9.8 µA, respectively. Beyond $B ≈$ 13 mT, the $I_c$ values decrease again upon further increasing the magnetic field. This double critical-current plateau effect in the magnetic-field dependence of $I_c$ is significantly different from the previously reported behavior of the critical current of microwires with one hole or with a triangular hole pattern, where only a single current plateau around $B =$ 0 was observed [5, 15]. We therefore conclude that microwires with different hole arrangements can react very differently in an external magnetic field.

We note here that the magnetic-field values at the observed plateaus do not coincide with vortex-matching field $B_m ≈$ 52 mT, which we estimated from $B_m ≈ \Phi_0/\Delta x^2$, assuming one magnetic-flux quantum per hole and geometric cell of area $\Delta x^2$. Although this fact may be counter-intuitive, it suggests that other factors primarily determine the occurrence and width of the $I_c(B)$ plateaus. Such a plateau has also been observed in lines with a single hole [15], for which a matching field cannot be defined, and the width of the resulting plateau was clearly related to the diameter of the hole. It has been argued that, in bridges without holes, vortices form at the edges. By contrast, in the plateau region of corresponding pinhole bridges, vortex-antivortex pairs are generated near the holes [15], which aligns with our observations (see Fig. 3(c,d) and the movies in the Supplementary material).

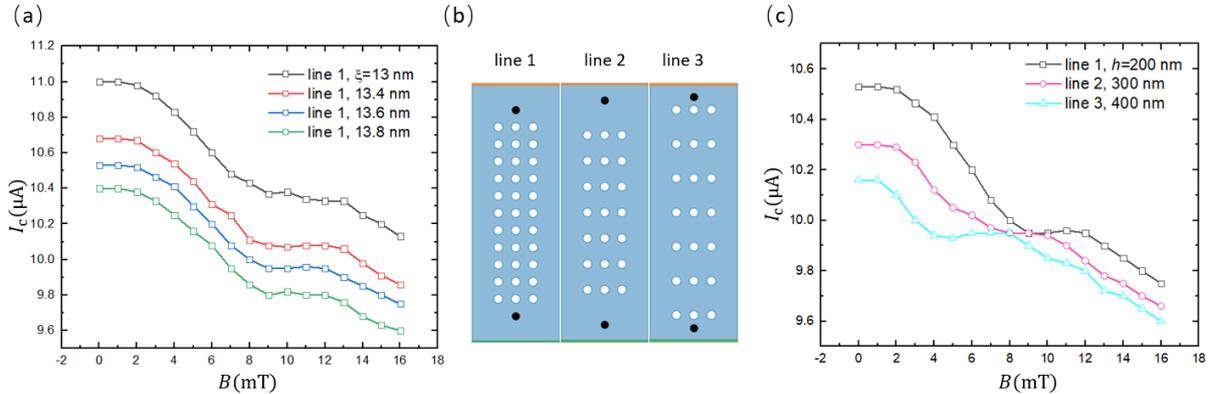

*Figure 4. (a) Simulated critical current vs. magnetic field of line 1 for four superconducting coherence lengths, $ξ$ = 13 nm (black), 13.4 nm (red), 13.6 nm (blue) and 13.8 nm (green); (b) hole arrangements for lines 1-3 with vertical distances $\Delta y$ =200 nm (line 1), 300 nm (line 2) and 400 nm (line 3) between the holes, with identical dimensions $3\times 1 µm^2$ of the superconducting film and pinhole diameter $D$ =100 nm. (c) simulated critical current vs. magnetic field for three lines with fixed coherence length $ξ$ = 13.6 nm and identical voltage threshold $4 V_0$ (≈ 9 mV).*

To study a possible effect of the longitudinal hole spacing, we performed additional simulations for microwires with vertical hole distances $\Delta y$ = 300 nm (line 2) and 400 nm (line 3), with the same transverse hole distance $\Delta x$ = 200 nm and hole diameter $D$ = 100 nm, patterned in bridges of identical size 3x1µm$^2$. As a result, the number of rows decreases from 11 (line 1) to 7 (lines 2 and 3). Figure 4 (c) shows the dependence of the critical current on magnetic field obtained for these microwires, fixing the coherence length to $\xi$ = 13.6 nm and using the same voltage threshold as for line 1. Two critical-current plateaus in $I_c(B)$ develop also in the lines 2 and 3. The critical currents for the first plateau ($B$ = 0) decrease systematically with increasing hole distance, from 10.53 µA to 10.3 µA and 10.16 µA, respectively. The $I_c$ value at the second current plateau is approximately 9.95 µA, independent of $\Delta y$, which may be a coincidence, however. By contrast, the magnetic-field range spanning the second plateau clearly and systematically shifts to smaller magnetic fields (from 9 ~12 mT in line 1 to 4 ~ 8 mT in line 3). As the number of pinhole rows is identical in lines 2 and 3, we may tentatively conclude that the observed tendency of an overall decrease of $I_c$ and shift of the second plateau to lower magnetic fields with increasing longitudinal hole spacing is an intrinsic trend, which is likely caused by a systematic variation of the respective vortex dynamics with varying hole distance [23-28]. It is therefore conceivable that in the hypothetical limit of an infinitely large $\Delta y$, corresponding to a single row of pinholes, only a single plateau remains around $B$ = 0, as it has been observed in experiments with a single defect [15].

## Experimental confirmation

In order to test the py-TDGL simulation results for superconducting micro-lines with pinholes with respect to the vortex dynamics, and in particular to the possible appearance of two critical-current plateaus in a magnetic field, we performed experiments on dedicated WSi microwires. Figure 5 shows the SEM images of three structures made of a 2 nm thick WSi film, with exact composition W$_{0.59}$ Si$_{0.41}$, sheet resistance 1000 Ω/sq, and a critical temperature of 2.4 K [5]. The dimensions of the microwires were $w$ = 1 µm and $L$ = 90 µm (Fig. 5 (a)). The pinhole arrangements and sizes in the experimentally investigated micro-lines (denoted in the following as exp-line 1, 2 and 3) are the same as the model lines in Fig. 1 and 4, with a transverse spacing $\Delta x$ = 200 nm and vertical spacings $\Delta y$ = 200 nm, 300 nm, and 400 nm, respectively, see Fig. 5(b,c,d). The measurements were conducted in a dilution refrigerator (Oxford Instruments Inc.) with a base temperature of 250 mK. The magnetic field was generated by a built-in commercial superconducting magnet and controlled by a Mercury iPS power system with an accuracy of 1 mT. For measuring critical currents and detecting voltage pulses we used a commercially available single-photon probing platform with an integrated low-noise voltage bias and amplifier (PHOTEC), photon counter SR400, and DPO 7354 Oscilloscope.

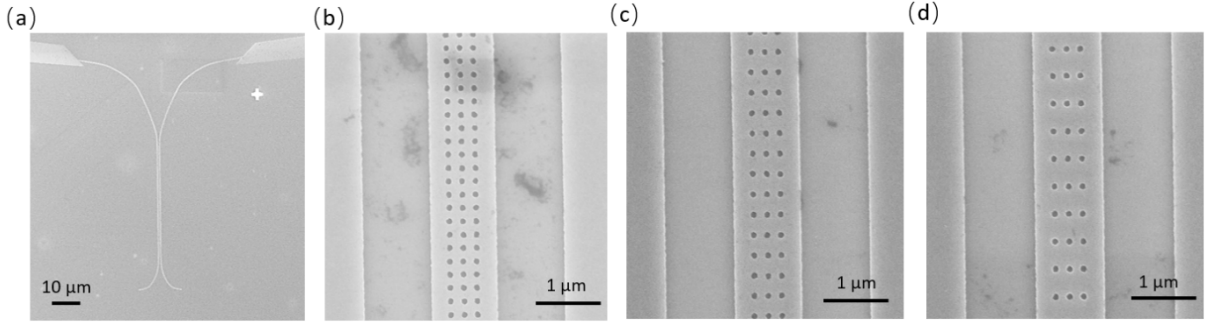

*Figure 5. SEM images of the samples used in the experiments: (a) 1 µm wide and 90 µm long straight microwire with a pinhole grid; (b) enlarged view of exp-line 1, with pinhole diameter D =100 nm, and transverse and longitudinal hole spacing hole $\Delta x = \Delta y$ = 200 nm, respectively; (c) exp-line 2 with $\Delta y$ = 300 nm; (d) exp-line 3 with $\Delta y$ = 400 nm.*

The magnetic-field dependence of the critical currents $I_c(B)$ of all three lines with pinholes are displayed in Fig. 6(a). They all show a comparable behavior with two plateaus of the critical current, see Fig. 6(b)). At the first plateau around zero magnetic field, the critical current values $I_c(0)$ are 10.65 µA, 10.25 µA, and 10.10 µA for the exp-lines 1-3, respectively. As predicted by the results of the simulations discussed above, we observe a second plateau in all three types of lines, with critical-current values 9.75 µA, 9.65 µA and 9.25 µA, respectively. These experimental results not only clearly confirm the occurrence of two critical-current plateaus for a rectangular pinhole pattern in superconducting bridges, but also correctly reproduce the qualitative correlation between the value of the critical current and the longitudinal hole spacing as predicted by the py-TDGL numerical simulations. In Fig. 6(c), we tentatively directly compare the experimental and simulated $I_c(B)$-dependences from Fig. 4(a) for exp-line 1, with the best match for $\xi$ = 13.4 nm and 13.6 nm.

We cannot expect a perfect reproduction of all experimental data by our simulations, however. While the second plateaus are well developed in the experiments, they are all located between approximately 6 mT and 14 mT and do not show any clear shift with varying $\Delta y$ as suggested by the simulated results shown in Fig. 4(c). Moreover, the measured magnetic-field dependences of $I_c(B)$ within and beyond the second plateaus are stronger than predicted [20]. Possible reasons for these discrepancies are manifold. First, the lines studied in the experiments were considerably longer than those in the simulations for practical reasons. Additional influences may be an inadequate modelling of the properties and geometry at the edges of the lines, the pinholes and the electrodes, and deviating material parameters. As we also disregarded possible thermal latching processes in the simulation due to the difficulty to model all relevant heat flows (see Supplementary material for more details), it is conceivable that heating effects and thermal latching are responsible for the steeper transition of the critical current from the first to the second plateau and the lower experimentally measured critical current at the second plateau, in comparison with the calculated $I_c(B)$ that do not include such thermal effects, as shown in Fig. 6(c).

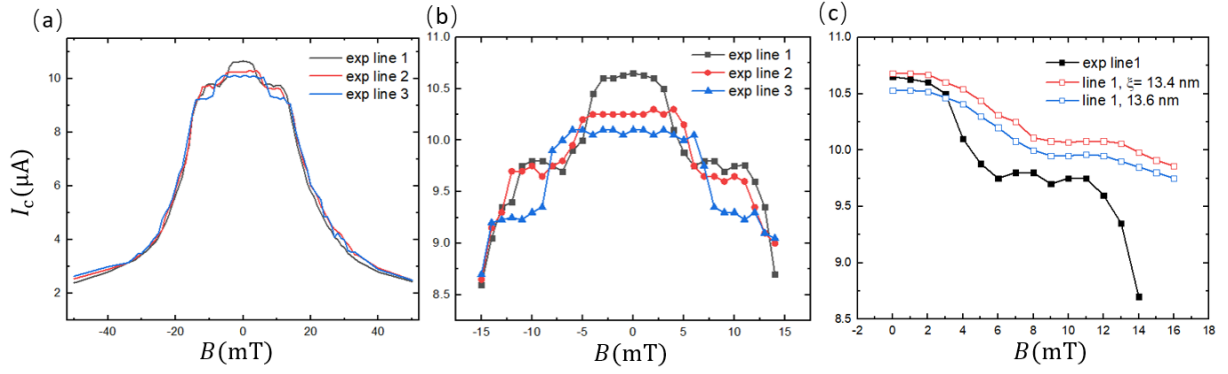

*Figure 6. (a) Magnetic-field dependence of the critical current $I_c$; black, red, and blue curves are for exp-lines 1, 2, and 3, respectively; (b) enlarged view of the same for magnetic fields between -15 mT to 15 mT; (c) comparison of the simulated $I_c(B)$ data with the experimental values for exp-line 1.*

Nevertheless, the still surprisingly good agreement between theory and experiment may be due to the fact that WSi is a weakly pinning material. Similar superconducting materials with comparable or even weaker bulk pinning are therefore preferable candidates for confirming the theoretical predictions of vortex dynamics in the absence of bulk pinning. Strong bulk pinning of vortices may effectively modify the properties of superconducting films with an artificial pinhole pattern, making the results of corresponding measurements incomparable to those obtained by theoretical models based on the assumption of zero bulk pinning.

## Conclusions

In summary, we have studied the vortex dynamics in superconducting microwires with perforated rectangular pinhole structures. The results of simulations based on the time-dependent Ginzburg-Landau equations are compared with corresponding experiments on the magnetic-field dependence of the critical current for three different longitudinal pinhole spacings. The simulations reveal three qualitatively different vortex states, namely a vortex free state in low currents below the critical current, and with increasing current a state with regular vortex flow, followed by a state with a more complex vortex pattern. The occurrence of two distinct plateaus in the field dependence of $I_c(B)$ contrasts with the observation of a single plateau around $B = 0$ in structures with a single hole or a triangular pinhole arrangement. This field dependence, as well as the qualitative dependence of $I_c$ on the vertical hole spacing was experimentally confirmed on WSi microwires with the identical hole arrangements, thereby confirming the validity and power of the numerical simulations. Certain deviations from the theoretical predictions, such as the observed faster decrease of $I_c(B)$ with increasing magnetic field beyond the second plateau and the exact location of the second plateau, may originate from an oversimplification of the geometrical structure, e.g., by implicitly assuming uniform edge pinning along the edges of the microwires and thus neglecting edge roughness and imperfections, or the disregard of a possible thermal latching process.

Despite these shortcomings, our approach may stimulate further related investigations, both theoretically and experimentally, on superconducting microwires containing artificial or natural defects for photon-detection applications.

## Acknowledgments

This work was supported by the Swiss National Foundation under Grant No. 20-175554. We thank to Dr. Logan Bishop-Van Horn for fruitful discussions.

## Data availability statement

All data that support the findings of this study are included within the article (and any supplementary files).

# Supplementary material for "Magnetic-field dependent vortex dynamics and critical currents in superconducting microwires with regular large-area perforation by pinholes"

## Kinetic inductance and London penetration depth

We show in Fig. S1(a) an SEM picture of the large inductor connected to the microwire exp-line 1 that we used for the kinetic- inductance measurements to estimate the London penetration depth $\lambda$. This large meander structure is composed of 5-micron wide wires consisting of approx. 5490 squares. The 1-micron wide exp-line (approx. 90 squares) described in the main article is located at the upper right end. Fig. 1S(b) shows a typical voltage pulse due to a dark-count event displayed on the oscilloscope with a reset time of 65 ± 5 ns, and thus the kinetic inductance per square is $L_K$ = 65 ns x 50 Ω /5580 sq = 582 pH/sq, with a resulting London penetration depth $\lambda = \sqrt{\frac{L_k d}{\mu_0}} \sim 960$ nm, where $d = 2$ nm is the thickness of WSi film.

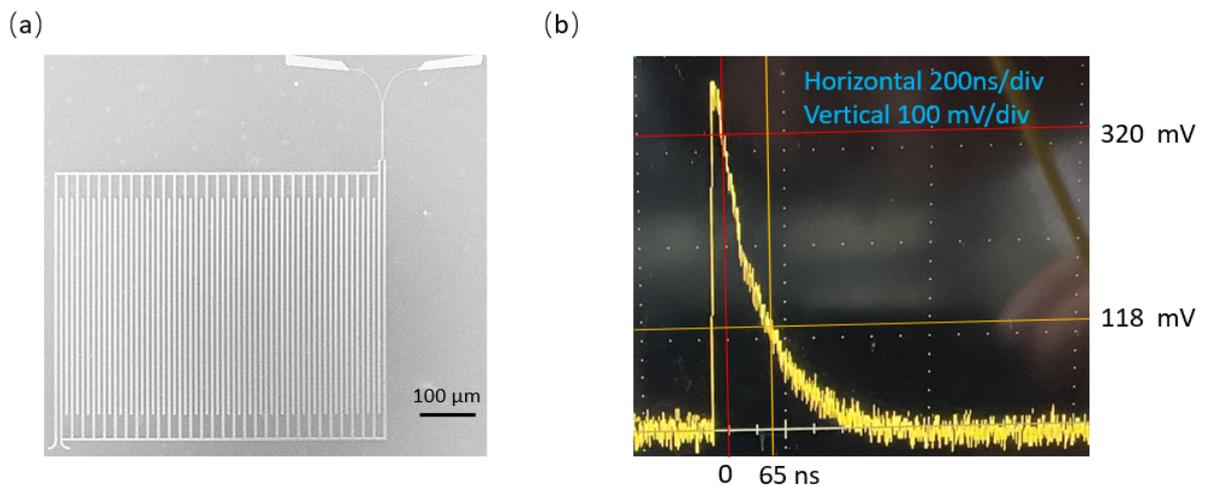

*Figure S1. (a) SEM image of the 5 µm wide large meander inductor connected to the exp-line 1 used for kinetic inductance measurement; (b) Voltage-pulse shape of one dark-count event measured for I = 10.5 µA and in B = 0 T, with a reset time of 65 ns as derived from the exponential decay of voltage.*

## Current-Voltage curves and critical current

We present in Fig. S2 the I-V curves of the model line-1 for a coherence length of 13.6 nm in different magnetic fields, from which we determined the critical-current values.

In low magnetic fields ranging from 0 to 2 mT (i.e., in the first critical-current plateau), the IV curves essentially overlap. Below approximately 10.5 µA, a small step of 2 voltage units is visible, which is comparable to the step-like effect in the voltage due to the discrete number of rows showing vortex flow near zero field, as shown in Fig. 3(a) and 3(c) in the main text. The first two such lines always appear first and simultaneously at both ends of pinhole arrangement, see Fig. 3(c) in the main text. In the corresponding experiments, however, the number of pinhole rows in the used samples by far exceeded that of the model samples for the simulation, so that corresponding voltage steps could not be resolved experimentally.

The voltage above 10.5 µA shows a steep rise, reflecting that the number of vortex-flow rows becomes sensitive to the bias current. Beyond 10.6 µA, the voltage shows a linear characteristic, and at all pinhole rows exhibit vortex-flow.

In medium magnetic fields, ≈ 5-7 mT, the IV curves gradually soften and vary continuously with the bias current. The curves for different magnetic fields are clearly separated, reflecting the decrease of the critical current. The IV curves above 10.6 µA again roughly coincide with the low-field data, indicating that the magnetic field has little effect on the vortex dynamics there. At higher magnetic fields (9 -11 mT, i.e., in the second plateau), the IV curve coincide again and show approximately the same linearity as with higher bias current, and all these IV curves are insensitive to magnetic field variations above 10.6 µA.

Do define a critical current, we chose a threshold value of 4 voltage units (see main text).

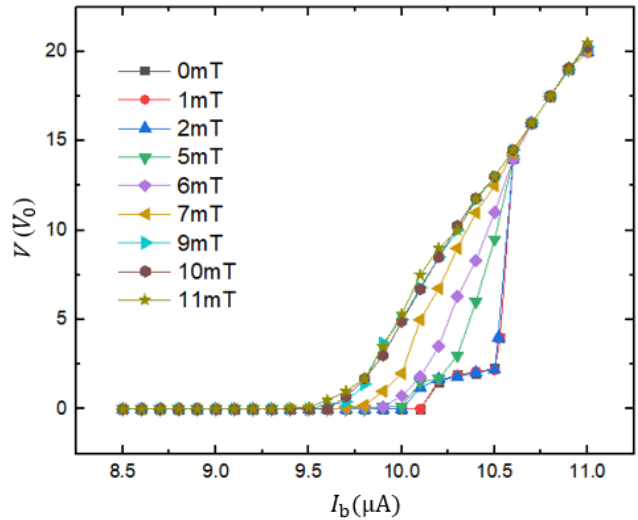

*Figure S2. Simulated I-V curves for the model micro-line 1 in low (0-2 mT), medium (5-7 mT) and high magnetic fields (9-11 mT), λ=960 nm, ξ=13.6 nm.*

## Heating and latching

We had disregarded possible thermal latching processes in the simulation due to difficulty to exactly model all relevant heat flows. As we stated in the main text, heating effects and thermal latching are possible reasons for the steeper transition of the critical current from the first to the second plateau and the lower experimentally measured critical current at the second plateau in comparison to the calculated $I_c(B)$. After vortex penetration, dissipative vortex motion and the associated generated voltage drop across the strip can lead to local heating due to non-ideal cooling and the formation of resistive or even fully normal areas in the strip. These normal areas are fully excluded from the carrying a supercurrent, thereby decreasing the remaining effective superconducting width of the strip. The size of these normal areas can therefore be effectively larger than what would be expected in the case of a perfect cooling of the device when the heat dissipated by the accelerated quasiparticles would immediately flow out to the substrate. The resulting increase of the size of the normal area can be understood as a result of the thermal flow from these heated areas (with an effective temperature above $T_c$) towards neighboring parts of the strip, thus increasing their effective temperature as well. Therefore, the measured critical current in such a process is lower than one would ideally expect. Thermal effects are stronger at low operating temperatures due to decrease of both thermal conductivities and heat capacities with decreasing temperature.

## Simulations

Animations of the time evolution of the order parameter, its phase, the supercurrent density and the normal-current density obtained from the simulations using the python computational package py-TDGL [1] for various values of magnetic field and bias current are available on a separate file (https://www.physik.uzh.ch/groups/schilling/paper/Timeevolution.pptx).

1. Horn, L.B.-V. 2023 pyTDGL: Time-dependent Ginzburg-Landau in Python *Computer Physics Communications* **291** 108799 https://doi.org/10.1016/j.cpc.2023.108799